\documentclass[11pt]{article}
\usepackage{amsmath}
\usepackage{amssymb}
\usepackage{cite}
\usepackage[cp1251]{inputenc}
\usepackage{amsfonts}
\usepackage{latexsym}
\usepackage{graphicx}
\usepackage{color}
\usepackage[english]{babel}
\usepackage{indentfirst}

\def\eq{\eqref}

\def\nb{\nabla}
\def\nn{\nonumber}
\def\cN{{\cal N}}
\def\cD{{\cal D}}
\def\cA{{\cal A}}
\def\cB{{\cal B}}
\def\cG{{\cal G}}
\def\cV{{\cal V}}
\def\cW{{\cal W}}
\def\cF{{\cal F}}
\def\bea{\begin{eqnarray}}
\def\tr{{\rm Tr }}
\def\beq{\begin{gather}}
\def\eea{\end{eqnarray}}
\def\eeq{\end{gather}}
\def\ln{\,\mbox{ln}\,}

\def\tr{\,\mbox{tr}\,}

\def\Tr{\,\mbox{Tr}\,}

\def\ve{\varepsilon}

\def\f{\frac}
\def\sB{\stackrel{\frown}{\square}}

\begin{document}

\begin{titlepage}

\begin{center}
\vspace{1cm} {\Large\bf One-loop divergences in the six-dimensional
$\cN=(1,0)$ hypermultiplet self-coupling model}

\vspace{1.5cm} {\bf A.S.
Budekhina\footnote{budekhina@tspu.edu.ru}$^{\,a,b}$, B.S.
Merzlikin\footnote{merzlikin@tspu.edu.ru}$^{\,c,a}$ }

\vspace{0.4cm}

{\it $^a$ Center for Theoretical Physics, Tomsk State Padagogical
University, \\
634061, Tomsk, Russia
\\ \vskip
0.15cm $^b$ National Research Tomsk State University, 634050, Tomsk,
Russia
\\ \vskip
0.15cm
$^c$  Tomsk State University of Control Systems and Radioelectronics,\\
634050, Tomsk, Russia\\
}
\end{center}
\vspace{0.4cm}
\begin{abstract}
We study the six-dimensional $\cN=(1,0)$ supersymmetric
hypermultiplet model with arbitrary self-coupling. The model is
considered in the external classical gauge superfield background.
Using the harmonic superspace formulation we study the one-loop
effective action of the model. We calculate the one-loop divergences
in the theory as a gauge-invariant function of the external gauge
multiplet. We demonstrate that the one-loop divergences in the
theory do not vanish even in the case of on-shell hypermultiplet
background. We briefly discuss an  application of the developed
technique to the hypermultiplet model in four dimensions.
\end{abstract}
\end{titlepage}

\setcounter{footnote}{0} \setcounter{page}{1}

\section{Introduction}

The study of the scalar field models in diverse dimensions attracted
much interest in the last few years. The $O(N)$ vector model of the
scalar fields with the relevant interactions in six dimensions has
been extensively studied in the context of a critical phenomenon and
large $N$ expansion \cite{FGK-14,FGKT-15} (see, also, the reviews
\cite{AGOR, Gr,MZ}). Supersymmetry restricts possible interaction of
the scalar fields, which are described in the general case by the
supersymmetric nonlinear sigma-models. These theories highlight the
close connections between supersymmetry and complex geometry. For
the case of the simplest supersymmetric sigma-model with four
supercharges the target space is known to be a K\"{a}hler manifold
\cite{Zum}. The rigid supersymmetric sigma-model with eight
supercharges in diverse dimensions possesses by huper-K\"{a}hler
manifolds as target space. Such sigma-model has $\cN = 2$
supersymmetry in four dimensions and $\cN=(1,0)$ supersymmetry in
the six ones \cite{AGF,ST-83}. In its turn, the target space of the
locally supersymmetric sigma-model with eight supercharges is
quaternionic K\"{a}hler manifold \cite{BW} (see, also,
\cite{Kuz12,HKLR} and references wherein).

The formulation of the supersymmetric sigma-models in terms of
unconstrained  superfields allows one to deeply understand the
relationship between geometry and supersymmetry. The superfield
technique is a convenient tool to study quantum properties of the
model, which preserves explicit supersymmetry during quantization
(see, e.g., \cite{GGRS})\footnote{See also \cite{BBP}, where the
quantum corrections for general four-dimensional supersymmetric
K\"{a}hler sigma models were studied in the superspace.}. In the
present paper, we study the six-dimensional $\cN=(1,0)$
supersymmetric hypermultiplet model minimally interacting with
external classical gauge multiplet. We consider the case of
arbitrary gauge-invariant potential function for the hypermultiplet
\cite{ST-84}. For the vanish gauge multiplet, this theory describes
the supersymmetric sigma-model with general multicenter
hyper-K\"{a}hler metric \cite{GIOS-85} (see, also, for a review
\cite{GIOS}).  We formulate the model in the six-dimensional
$\cN=(1,0)$ harmonic superspace \cite{GIOS,GIKOS,HSW,Z} and develop
the covariant technique to study the effective action. We construct
the one-loop effective action in the theory using  the proper-time
technique in the harmonic superspace \cite{BIMS-a,BIMS-b,BBKO,BK}.

The study of the ultraviolet behavior of supersymmetric field
theories in six dimensions has received considerable attention in
recent years. The analysis of the UV divergences is carried out
using both symmetry considerations \cite{HS2,ISZ,BHS1,BHS2,BIS} and
direct calculations of the effective action
\cite{Kazakov,BP,BIMS-a,BIMS-b,BIMS-c,BBM,CT} (see also earlier
analysis \cite{FT, MS1,MS2,HS1}) and scattering amplitudes
\cite{Amplitudes1,Amplitudes2,Amplitudes3,Amplitudes4,Amplitudes5}.
As an application of the developed technique, we calculate the
one-loop divergent contribution to the effective action in the
gauge-invariant $\cN=(1,0)$ supersymmetric sigma-model in six
dimensions \cite{ST-84}. The obtaining result depends on both
background hypermultiplet and external gauge multiplet.  We also
demonstrate that the leading finite contribution to the effective
potential of the scalar fields in the model vanishes.

The work is organized as follows. In Section 2 we discuss the basic
properties of six-dimensional $\cN=(1,0)$ harmonic superspace and
formulate the model. Section 3 contains the quantization procedure
in  the $\cN=(1,0)$ harmonic superspace and the evaluation  of the
one-loop effective action for the model. Analysis of divergent
contributions to the one-loop effective action  is  collected in
Section 4. Finally, in Section 5, we briefly discuss the application
of the developed technique to the $\cN=2$ supersymmetric
hypermultiplet model in four dimensions.

\section{Hypermultiplet theory in $6D$}

Throughout out the paper, we follow the notations and conventions
from the work \cite{ISZ}. We denote the $6D$ Minkowski space
coordinates by $x^M$ and the Grassmann ones by $\theta^{a}_i$, where
$M=0,\ldots,5$,\ $a=1,\ldots, 4$. The additional $SU(2)$ index
$i=1,2$ corresponds to the R-symmetry group of the simplest ${\cal
N}=(1,0)$ supersymmetry in six dimensions.

We use the harmonic superspace formulation \cite{GIOS} for the
theory. The six-dimensional $\cN=(1,0)$ harmonic superspace
coordinates $(x^M,\theta^{ai}, u^{\pm i})$ include additional
harmonic variables $u^{\pm i}$. They parameterize the coset
$SU(2)/U(1)$ and obey the constraints $u^{+i} u_i^- = 1$, $u_i^-
\equiv (u^{+i})^*$. The harmonic superspace naturally admits the
analytic basis, which is parameterized by the coordinates
$\zeta=(x_{\cal A}^M, \theta^{\pm a})$
 \bea
x^M_{\cal A}=x^M + \frac i2  \theta^{+a}\gamma^M_{ab}\theta^{-b},
\quad \theta^{\pm a}= u^\pm_k\theta^{ak}.
 \eea

In the analytic basis the spinor and harmonic derivatives are
defined as follows \cite{ISZ}
 \bea
&& D^+_a=\partial_{-a}~, \quad D^-_a=-\partial_{+a} -2i\theta^{-b}\partial_{ab}\,,  \\
&& D^0 = u^{+i} \partial_{-i} - u^{-i} \partial_{+i} + \theta^{+a}
\partial_{+ a} - \theta^{-a} \partial_{-a}\,, \quad
D^{\pm\pm}=\partial^{\pm\pm}+i\theta^{\pm a}\theta^{\pm
b}\partial_{ab} +\theta^{\pm a}\partial_{\mp a}.
 \eea
Here we denoted $\partial_{\pm a}\theta^{\mp b} = \delta^b_a $,
$\partial^{\pm\pm} = u^{\pm i} \partial_{\pm i}$, $\partial_{\pm
i}=\frac{\partial }{\partial u^{\mp i}}$ and
$\partial_{ab}=\tfrac12(\gamma^M)_{ab}\partial_M$. The
six-dimensional Weyl matrices $(\gamma^M)_{ab}$ are chosen as
 \bea
(\gamma^M)_{ab}=-(\gamma^M)_{ba}\,\quad\tilde\gamma_M^{ab}=\tfrac12\varepsilon^{abcd}
(\gamma_M)_{cd}\,, \quad
(\gamma_M)_{ac}(\tilde\gamma_N)^{cb}+(\gamma_N)_{ac}(\tilde\gamma_M)^{cb}
=-2\delta^b_a \eta_{MN}\,,
 \eea
where $\varepsilon^{abcd}$ is the totally antisymmetric symbol. The
derivatives listed above satisfy the algebra
 \bea
&& \{D^+_a,D^-_b\}=2i\partial_{ab}, \quad [D^{++}, D^{--}] = D^0\,,
\quad  [D^{\pm\pm},D^\pm_a]=0\,, \quad [D^{\pm\pm},D^\mp_a]=D^\pm_a.
\label{alg1}
 \eea

We include the integration over harmonic variables $u^{+i}$ into the
full and analytic superspace integration measures,
 \bea
d^{14} z = d^6 x_{\cal A}du (D^{+})^4 (D^{-})^4\,, \quad
d\zeta^{(-4)} = d^6 x_{\cal A} du(D^{-})^4\,,
 \eea
where $(D^\pm)^4=-\tfrac{1}{24}\ve^{abcd}D^\pm_a D^\pm_b D^\pm_c
D^\pm_d$.

We consider the $\cN=(1,0)$ $q$-hypermultiplet model in six
dimensions. We assume the standard kinetic term and general
self-coupling potential for the superfield $q^+$ \cite{GIOS}. The
action for the model is
 \bea
S[q^+]=-\int d\zeta^{(-4)}\left[\widetilde q^{+}D^{++}q^{+} +
L^{+4}(q^+,\widetilde{q}^{+},u)\right], \label{S}
 \eea
where $L^{+4}$ is an arbitrary analytic function of the
hypermultiplet $q^+$, which is real one $\widetilde L^{+4} =
L^{+4}$. The kinetic term is also real under tilde conjugation
\cite{GIOS} because of the property $\widetilde{\tilde q^+}= -q^+$.
In the simplest quartic interaction, then
$L^{+4}=\tfrac{1}{2}(\widetilde{q}^+ q^+)^2$, the action \eqref{S}
describes the Taub-NUT sigma-model for the physical bosons (see,
e.g., \cite{GIOS} for details).

Let us introduce an abelian analytic gauge superfield $V^{++}$,
which is transformed as
 \bea
\delta V^{++} = - D^{++}\lambda\,, \label{trans1}
 \eea
where $\lambda = \lambda(\zeta,u)$ is a real analytic gauge
parameter. We assume the superfield $V^{++}$ is an external
background superfield minimally interacting with hypermultiplet. The
action describing this model is similar to \eqref{S} and can be
written as
 \bea
S[q^+;V^{++}] = -\int d\zeta^{(-4)}\left[\widetilde
q^{+}(D^{++}+iV^{++})q^{+} + L^{+4}(\widetilde q^{+}q^{+})\right]\,.
\label{S1}
 \eea
The transformations
 \bea
\delta q^{+} = i\lambda q^+\,, \quad \delta \tilde q^{+} = -i\lambda
\tilde q^+\,. \label{trans2}
 \eea
together with \eqref{trans1} hold the action \eqref{S1} invariant if
the potential function for hypermultiplet depends on the invariant
combination of superfields, $L^{+4}=
L^{+4}(\widetilde{q}^{+}q^{+})$. The action \eqref{S1} can be
considered as a part of a more general theory describing the
supersymmetric Yang-Mills theory in the broken phase interacting
with the self-coupling hypermultiplet.

Finally, it is interesting to discuss a case of $4N_f$
hypermultiplets $q^+_{{\cal A} I}$. Here we have denoted $q^+_{\cA
I} = (q^+_{1}...q^+_{N_f}, -\tilde q^+_1...-\tilde q^+_{N_f})_I$ for
each index $I=1,2$. Also,  the conjugation of the hypermultiplet
$q^+_{{\cal A} I}$ reads $\widetilde{q^+_{\cA I}} = q^{+\cA}_I =
\Omega^{\cal A B} q^+_{\cB I} $, where $\Omega^{\cal AB} =
-\Omega^{\cal BA}$ is the $Sp(N_f)$ invariant tensor. We consider
the model with the classical action
 \bea
S^{N_f}[q^{+};V^{++}] = -\int d \zeta^{(-4)} \Big( q^{+\cA}_I D^{++}
q^+_{\cA I} + V^{++} \epsilon_{IJ}\, q^{+\cA}_I q^+_{\cA J}  +
V^{++}\xi^{++}\Big)\,, \label{Sn}
 \eea
where the summations over indexes $\cA$ and $I$ are assumed and
$\epsilon_{12}= -\epsilon_{21} =1$. The last term in \eqref{Sn}
contains the constant vector $\xi^{++} = \xi^{ij} u^+_i u^+_j$,
which is under the constraint $D^{++}\xi^{++}=0$. This contribution
corresponds to the Fayet-Iliopoulos term in the  harmonic superspace
\cite{GIOS}. The action \eqref{Sn} is invariant under the local
$SO(2)$ symmetry
 \bea
 \delta V^{++} = -D^{++} \lambda, \qquad \delta {q}^+_{\cA I} = \lambda \epsilon_{IJ} q^+_{\cA J}\,,
 \eea
 where $\lambda = \lambda(\zeta,u)$ is a real analytic gauge parameter.

The bosonic part of the action \eq{Sn} has the form
 \bea
S^{N_f}_{\rm bos} &=&\int d^6 x \Big(\partial^{ab} \phi^{i \cA}_I \,
\partial_{ab} \phi_{i \cA I} - i A^{ab} \epsilon_{IJ}( \phi^{i \cA}_I\,
\partial_{ab} \phi_{i \cA J}-\partial_{ab}\phi^{i \cA}_I\, \phi_{i\cA J}) \nn \\
&&\qquad \qquad - A^{ab}A_{ab}\,\, \phi^{i\cA}_I \phi_{i\cA I} +
D_{ij}  (\epsilon_{IJ}\phi^{i\cA}_I \,\phi^j_{\cA J}+\xi^{ij})
\Big)\,, \label{Sn_bos}
 \eea
where $\phi^{i}_{\cA I}$ is the complex scalar component of
hypermultiplet $q^{+}_{\cA I}$ and $\phi_{i \cA I} = \epsilon_{ij}
\phi^{j}_{\cA I},\, \epsilon_{ij} =-\epsilon_{ji}$. The bosonic
components of the vector multiplet $V^{++}$ in the Wess-Zumino gauge
include the gauge field $A_{ab}$  and the triplet of auxiliary
scalar fields $D_{ij}$ \cite{ISZ}. The action \eq{Sn_bos} takes a
simpler form if we introduce the covariant derivative $\nb_{ab} =
\partial_{ab}-iA_{ab}$
 \bea
S^{\cN_f}_{\rm bos} = \int d^6 x \Big( \nb^{ab} \phi^{i \cA}_I
\nb_{ab} \phi_{i \cA I}  + D_{ij}  (\epsilon_{IJ}\phi^{i\cA}_I
\,\phi^j_{\cA J}+\xi^{ij})\Big)\,. \label{Sn_bos2}
 \eea
The fields $A_{ab}$ and $D_{ij}$ in the action \eqref{Sn_bos2}  one
can exclude using the algebraic equations of motion. In this case,
the equation for the field $D_{ij}$ can be considered as a
constraint on the dynamical   field $\phi^i_{\cA I}$.

\section{One-loop effective action}

To study the effective action in model \eq{S1} we will use the loop
expansion at leading order. Setting
 \bea
q^+ \rightarrow q^+ + Q^+\,, \qquad \widetilde q^+ \rightarrow
\tilde q^+ +\widetilde Q^+\,, \label{split}
 \eea
where $Q^+$ and $\widetilde Q^+$ are classical fields, we decompose
the classical action up to the second order over $q^+$ and $\tilde
q^+$
 \bea
S_{2} = - \int d\zeta^{(-4)}\Big\{ \tilde q^+ \cD^{++} q^+ +
\tfrac12 \Psi \Big((\tilde Q^+ q^+)^2 + 2 \tilde Q^+ Q^+ \tilde q^+
q^+  +(\tilde q^+ Q^+ )^2 \Big)\Big\}\,. \label{S2}
 \eea
Here we introduce the covariant harmonic derivative
 \bea
\cD^{++}=D^{++}+ i V^{++} + i \Psi^{++} \label{D}
 \eea
and use the notations
 \bea
\Psi^{++} = -i \f{\partial L^{+4}(\tilde Q^+ Q^+)}{\partial(\tilde
Q^+ Q^+)}\,, \quad \Psi = \f{\partial^2 L^{+4}(\tilde Q^+
Q^+)}{\partial(\tilde Q^+ Q^+)^2}\,. \label{Psi}
 \eea

We rewrite the action \eqref{S2} into the matrix form
 \bea
S_{2} = -\f12\int d \zeta^{(-4)} \widetilde{
    \left(\begin{array}{cc} q^+ & \tilde q^+ \\
        \end{array}
        \right)}
    \left(
    \begin{array}{cc}
        \cD^{++} + \Psi \tilde Q^+ Q^+ & \Psi (Q^+)^2 \\
        -\Psi (\tilde Q^+)^2 & \cD^{++} - \Psi \tilde Q^+ Q^+  \\
    \end{array}
    \right)
    \left(
    \begin{array}{c}
        q^+\\
        \tilde q^+ \\
    \end{array}
    \right)\,. \label{S3}
 \eea
Using the last expression it is easy to obtain the one-loop
contribution $\Gamma^{(1)}$ to the effective action of the theory
\eqref{S1} by integrating over quantum fields. We have
 \bea
\Gamma^{(1)}[Q^+;V^{++}] = \tfrac{i}2 \Tr_{(3,1)} \ln
S''_2[Q^+;V^{++}]\,. \label{G}
 \eea
In the \eqref{G} the functional trace includes the matrix trace and
integration over harmonic superspace
$$
\Tr_{(q,4-q)} {\cal O} = \tr \int d \zeta_1^{(-4)}d \zeta_2^{(-4)}
\, \delta_{\cal A}^{(q,4-q)}(1|2)\,  {\cal O}^{(q,4-q)}(1|2),
$$
where $ \delta_{\cal A}^{(q,4-q)}(1|2)$ is an analytic
delta-function \cite{GIOS} and  ${\cal O}^{(q,4-q)}$ is the kernel
of an operator  ${\cal O}$ acting in the space of analytic
superfields with the harmonic U(1) charge $q$ \cite{GIOS}. The
expression \eqref{G} includes the second variation derivative of the
action $S_2$ \eqref{S3} and can be represented in the form
 \bea
\Gamma^{(1)}[Q^+;V^{++}] &=& i \Tr_{(3,1)} \ln \cD^{++} +
\tfrac{i}2\Tr_{(3,1)} \ln\Big({ \bf 1} + \cG^{(1,1)} \Psi {\bf {\cal
Q}^{++}}\Big). \label{G1}
 \eea
Here we introduce the matrix
 \bea
{\bf {\cal Q}^{++}} = \left(
\begin{array}{cc}
        \tilde Q^+ Q^+  & ( Q^+)^2  \\
        -(\tilde Q^+)^2 & -\tilde Q^+ Q^+  \\
    \end{array}
    \right) \label{Q}
 \eea
and the Green function $\cG^{(1,1)}$, which  satisfies the equation
 \bea
 \cD^{++}_1\cG^{(1,1)}(1|2) = \delta^{(3,1)}(1|2)\,.
    \label{Gr}
 \eea

We consider the second term in the one-loop contribution to
effective action \eqref{G1} in more detail. This part of effective
action is defined as a series
 \bea
\Tr_{(3,1)} \ln\Big({ \bf 1} + \cG^{(1,1)} \Psi {\bf {\cal Q}^{++}}
\Big) = \Tr \sum_{n=1}^{\infty} \frac{(-1)^n}{n} \Big(\cG^{(1,1)}
\Psi {\bf {\cal Q}^{++}}  \Big)^n\,. \label{G2}
 \eea
The matrix power in the last expression also includes the
integration over analytic subspace. Every term in the series
\eqref{G2} including $({\bf {\cal Q}^{++}})^n = {\bf {\cal
Q}^{++}}_1 {\bf {\cal Q}^{++}}_2 \ldots {\bf {\cal Q}^{++}}_{n-1}
{\bf {\cal Q}^{++}}_1 $, $n>1$, contains the second power of the
matrix  ${\bf {\cal Q}^{++}}$ depending on the same argument  due to
the trace $\Tr$. However the matrix $ {\bf {\cal Q}^{++}}$ is a
traceless and a nilpotent one, $ ({\bf {\cal Q}^{++}})^2 = 0$. Hence
the contribution \eqref{G2} equals zero.

Thus the one-loop effective action \eqref{G1} is reduced to
 \bea
\Gamma^{(1)}[Q^+;V^{++}] &=& i \Tr_{(3,1)} \ln \cD^{++}\,.
\label{G3}
 \eea
We have to note that the effective action \eqref{G3} depends on the
two superfields, namely the external classical analytic gauge
superfield $V^{++}$ and the background hypermultiplet $Q^+$. We do
not assume any restriction on these fields. Our aim now is to study
one-loop divergencies of the action \eqref{G3}, but first of all, we
consider the properties of covariant harmonic derivative $\cD^{++}$.

The operator $\cD^{++}$ includes beside the analytic gauge
connection $V^{++}$ also the superfield $\Psi^{++}$ \eqref{Psi}. The
analytic superfield $\Psi^{++}$ by construction is an ordinary  real
function of the hypermultiplet. One can consider the superfield
$\Psi^{++}$ as an additional analytic gauge connection. Hence we
have the harmonic covariant derivative depending on the two abelian
gauge connections.

As a next step, we have to derive the algebra of covariant
derivatives including both gauge superfield connections. For
convenience, we introduce a notation
 \bea
\cV^{++} = V^{++} + \Psi^{++}.
 \eea
Then following standard procedure \cite{GIOS}  we consider the
non-analytic gauge connection $\cV^{--}$ by the rule
 \bea
D^{--}\cV^{++} = D^{++}\cV^{--}\,,
 \eea
which in our abelian case can be solved exactly
 \bea
\cV^{--} = \int d u_1 \frac{\cV^{++}(u_1)}{(u^+ u^+_1)^2}\,.
 \eea
The commutation relations of the algebra \eqref{alg1} should be
rewritten and up to the definitions coincide with the algebra of
covariant derivatives of six-dimensional $\cN=(1,0)$ supersymmetric
gauge theory \cite{ISZ,BIS}
 \bea
&& [\nabla^{--}, D^+_a] = \nabla^{-}_a\,, \quad [\nabla^{++},
\nabla^{-}_a] = D^+_a\,, \quad [\nabla^{++}, D^{+}_a] =
[\nabla^{--}, \nabla^{-}_a] =
0\,,\nonumber\\
&& [D^+_a, \nabla^-_b] = 2i \nabla_{ab}\,,\quad [D^+_a, \nabla_{bc}]
=\tfrac{i}2\varepsilon_{abcd} \cW^{+d}\,, \quad [\nabla^-_a,
\nabla_{bc}] = \tfrac{i}{2}\varepsilon_{abcd} \cW^{-d}\,.
\label{alg2}
 \eea
Here we have introduced the gauge field strength $\cW^{+a}$
 \bea
\cW^{+a}=-\tfrac{i}{6}\varepsilon^{abcd}D^+_b D^+_c D^+_d\cV^{--}\,,
\quad \cW^{-a}:= \nabla^{--}\cW^{+a}\,. \label{W}
 \eea

Let us define the analytic superfield ${\cal F}^{++} = D^+_a
\cW^{+a}$. The superfields $\cF^{++}$ and $\cW^{+a}$ possess some
useful properties \cite{BIS}
 \bea
D^{++} \cF = D^{++} \cW^{+a} = D^{++} \cW^{-a}= 0\,, \quad
\cW^{-a}=D^{--} \cW^{+a}\,. \label{pr}
 \eea
We have to note that superfields $\cF^{++}$ and $\cW^{+a}$ depend on
the superfields $V^{++}$ and $Q^+$. One can  separate both
$\cF^{++}$ and $\cW^{+a}$ in two independent parts
 \bea
\cW^{+a} = \cW^{+a}_V + \cW^{+a}_Q\,, \quad \cF^{++} = \cF^{++}_V +
\cF^{++}_Q\,. \label{F}
 \eea
Each of these superfields corresponds to the connection $V^{++}$ and
$\Psi^{++}$ respectively.

The covariant analytic d'Alembertian $\sB$ \cite{BIS} in six
dimensions reads
\begin{equation}\label{smile}
\sB=\tfrac{1}{2}(D^+)^4(\cD^{--})^2\,.
\end{equation}
Acting in a space of analytic superfields the operator \eq{smile}
transforms to \cite{ISZ}
 \begin{eqnarray}
\label{Box_First_Part} \sB = \eta^{MN} \cD_M \cD_N + \cW^{+a}
\cD^{-}_a + \cF^{++} \cD^{--} - \tfrac{1}{2}(\cD^{--} \cF^{++}),
    \end{eqnarray}
where $\eta_{MN}$ is $6D$ Minkowski metric and $\cD_M =\partial_M -
i {\cal A}_M$ is a space-time covariant derivative, $\cA_M =
\tfrac{i}2 (\tilde\gamma_M)^{ab} D^+_a D^+_b \cV^{--}$. Also we have
introduced    the covariant harmonic derivative $\cD^{--}=D^{--}+ i
\cV^{--}$,    which is constructed using non-analytic connection
$\cV^{--}$.

\section{One-loop divergencies}

The one-loop contribution to effective action \eqref{G3} contains
the first order differential operator $\cD^{++} = D^{++} +
i\cV^{++}$. To calculate the divergent contributions we following
\cite{BK} and  vary the one-loop effective action \eq{G3} with
respect to the superfield  $\cV^{++}$
 \bea
\delta \Gamma^{(1)} = - \Tr_{(3,1)} \delta\cV^{++} \cG^{(1,1)} = -
\int d\zeta_1^{(-4)} \delta\cV_1^{++} \cG^{(1,1)}(1|2)\Big|_{2\to 1
}\,. \label{vG}
 \eea
The Green function $\cG^{(1,1)}(1|2)$ solves the equation \eq{Gr}
and has the following formal expression (see, e.g., \cite{BIMS-a},
\cite{BP})
 \bea
\cG^{(1,1)}(1|2) = \f{(D_1^{+})^4 (D_2^+)^4}{\sB_1}
\f{\delta^{14}(z_1-z_2)}{(u^+_1 u^+_2)^3}\,, \label{Gr2}
 \eea
where $\delta^{14}(z_1-z_2) =
\delta^6(x_1-x_2)\delta^{8}(\theta_1-\theta_2)$ is a full
$\cN=(1,0)$ superspace delta-function.

The analysis of divergent contributions of the expression \eq{vG} is
similar to what was carried out for six-dimensional $\cN=(1,0)$
supersymmetric abelian gauge theory in \cite{BIMS-a}. First of all
we use the proper time technique for the inverse operator
$\sB{}^{-1}$ and rewrite the expression \eq{vG} as follows:
 \bea
\delta \Gamma^{(1)} =-\int d\zeta_{1}^{(-4)}\delta \cV^{++}
\int_0^\infty d(is)(is\mu^2)^{\f\varepsilon2}
e^{is\sB_1}(D_1^+)^4(D^+_2)^4
\f{\delta^{14}(z_1-z_2)}{(u^+_1u^+_2)^3}\Big|^{2=1}. \label{G4}
 \eea
In the last expression, we have introduced the proper-time parameter
$s$ and the arbitrary parameter $\mu$ of mass dimension. We use a
dimensional regularization scheme, where divergencies appear as a
pole $\tfrac1\varepsilon$ under the condition $\varepsilon \to 0$.

The integrand in the expression \eq{G4} contains eight $D$-factors
acting on the Grassmann delta-function $\delta^8(\theta_1-\theta_2)$
in coinciding points limit. To calculate this limit one can use the
identity
$$(D_1^+)^4(D^+_2)^4 \delta^8(\theta_1-\theta_2)|_{2\to 1} =
(u^+_1u^+_2)^4.$$ After that, it transforms to
 \bea
\delta \Gamma^{(1)}_{\rm div}= -\int d\zeta_{1}^{(-4)}\delta\cV^{++}
\int_0^\infty d(is)(is\mu^2)^{\f\varepsilon2}
e^{is\sB_1}(u^+_1u^+_2) \delta^{6}(x_1-x_2)\Big|^{2=1}_{\rm div}.
\label{G5}
 \eea
Operator $\sB$ contains the harmonic derivative $D^{--}$ which
acting on the factor $(u^+_1u^+_2)$ produces non-zero contribution
due to the identity $D_1^{--}(u^+_1u^+_2)|_{2\to 1}  =
(u^-_1u^+_2)|_{2\to 1} =-1$. Commuting the operator $e^{is\sB}$ with
the factor $(u^+_1u^+_2)$ we act on the space-time delta-function
and count integral over proper time (see the details in
\cite{BIMS-a}). We have
 \bea
\delta \Gamma^{(1)}_{\rm div} =  \f{1}{3 (4\pi)^3 \varepsilon}\int
d\zeta^{(-4)} \,\delta \cV^{++}\, \partial^2 \cF^{++}\,,
 \eea
where $\partial^2 = \partial_M\partial^M = \tfrac12
(D^{+})^4(D^{--})^2$. One can show that the last expression can be
evaluated from the functional
 \bea
\Gamma^{(1)}_{\rm div} =  \f{1}{6 (4\pi)^3 \varepsilon}\int
d\zeta^{(-4)} (\cF^{++})^2\,, \label{G6}
 \eea
taking into account the connection of the variation $\delta
\cV^{++}$ with $\delta \cV^{--}$ \cite{ISZ}
 \bea
\delta \cV^{--} = \tfrac12(D^{--})^2 \delta \cV^{++} - \tfrac12
D^{++}(D^{--}\delta \cV^{--})\,,
 \eea
and the properties \eqref{pr} for the superfield $\cF^{++}$.

The one-loop divergent contribution \eq{G6} to the effective action
of the theory \eqref{S1} depends on the external gauge multiplet
$V^{++}$ and background hypermultiplet $Q^+$.  Using the definitions
\eq{F} we can rewrite \eq{G6} in the form
 \bea
\Gamma^{(1)}_{\rm div}[V^{++},Q^+] =  \f{1}{6 (4\pi)^3
\varepsilon}\int d\zeta^{(-4)} (\cF^{++}_V+\cF^{++}_Q)^2\,.
\label{G7}
 \eea

Let us discuss the final result \eq{G7}.  First, we consider
possible constraints on the background superfields. The gauge
superfield $V^{++}$ is an arbitrary external classical superfield.
One can restrict the superfield $V^{++}$ requiring, e.g.,
 \bea
\cF^{++}_V = 0\,. \label{con1}
 \eea
This condition  corresponds to the Maxwell equation for the physical
boson component of superfield $V^{++}$.  If we assume that external
superfield $V^{++}$ satisfies the constraint \eq{con1} the divergent
contribution will depend only on  off-shell background superfield
$Q^{+}$.  Hence, the one-loop divergent contribution \eq{G7} under
the condition \eq{con1} can be rewritten in the form \bea
\Gamma^{(1)}_{\rm div}[Q^+] = \f{1}{6 (4\pi)^3 \varepsilon}\int
d\zeta^{(-4)} (\cF^{++}_Q)^2 = \f{1}{6 (4\pi)^3 \varepsilon}\int
d^{14} z \Psi^{--}\partial^2\Psi^{++}. \label{G8} \eea Here we use
the properties $\cF^{++} = \tfrac12 D^{++} D^{--}\cF^{++}$ and
$D^{++}\cF^{++}=0$.

The superfield  $\cF^{++}_Q$ depends on the dynamical classical
background hypermultiplet $Q^{+}$. The classical equation of motion
for the background hypermultiplet
 \bea
\cD^{++} Q^{+} = (D^{++} + i V^{++} + i \Psi^{++})Q^+= 0\,,
\label{con2}
 \eea
includes self-interaction term $\Psi^{++}$  and external gauge
multiplet $V^{++}$. This equation also should be completed by the
equation for Grassmann conjugate hypermultiplet $\tilde Q^+$.
Equation \eq{con2} gives the nontrivial connection between physical
components of hypermultiplet $Q^+$ and external gauge multiplet
$V^{++}$ \cite{GIOS}. If we take into account condition \eq{con2},
divergent contribution \eq{G8} will contain the dependence on the
external gauge multiplet in the component action.

As an example, we consider the case when external gauge multiplet
$V^{++}=0$ and the potential function $L^{+4}$ in the initial model
\eq{S1} has a simple form $L^{+4}= \tfrac{1}{2}(\tilde Q^+ Q^+)^2$.
The analytic $\Psi^{++}$ and non-analytic $\Psi^{--}$ connections
for this potential are
 \bea
\Psi^{++} = -i\tilde Q^+ Q^+\,, \quad \Psi^{--} = \int d u_2
\frac{\Psi^{++}_2}{(u_1^+u_2^+)^2}\,.
 \eea
The divergent contribution \eq{G8} in the theory with such
self-interaction takes the form
 \bea
\Gamma^{(1)}_{\rm div}[Q^+] = -\f{1}{6 (4\pi)^3 \varepsilon}\int
d^{14} z \f{du_1du_2}{(u_1^+u^+_2)^2}(\tilde Q^+
Q^+)_1\partial^2(\tilde Q^+ Q^+)_2\,. \label{G9}
 \eea
The last expression is non-local in the harmonic space. However for
the on-shell background hypermultiplet the non-locality of the
divergent contribution can be removed. In this case, $\Psi^{--} = -i
\tilde Q^- Q^-$, where $Q^-=D^{--}Q^+$, and we finally have
 \bea
\Gamma^{(1)}_{\rm div}[Q^+] = -\f{1}{6 (4\pi)^3 \varepsilon}\int
d^{14} z\, \tilde Q^- Q^-\partial^2(\tilde Q^+ Q^+)\,. \label{G10}
 \eea

As one can see the divergent contribution \eqref{G10} vanishes if we
suppose the slowly varying background hypermultiplet, $\partial Q^+
\approx 0$. To obtain the leading  finite contribution to effective
action \eqref{G4} in this case one can use the method  developed in
\cite{BMP}. The result is
 \bea
\Gamma^{(1)}_{\rm lead}[Q^+] \sim \int d\zeta^{(-4)} ({\cal
W}_Q^+)^4 = 0\,.
 \eea
Indeed, the superfield ${\cal W}^{+ a}_Q$ introduced in \eqref{W}
includes the term $D^+_a D^+_b Q^-$, which after some algebra
transforms to $2i \partial_{ab} Q^+$ and vanishes for the constant
background. We can consider this as a consequence of the absence of
a scalar  potential in the gauge-invariant supersymmetric
$\sigma$-model in six dimensions \cite{ST-84}.

The developed technique in application to the model \eq{Sn} takes
the following result
 \bea
\Gamma^{(1)\, N_f}_{\rm div}[V^{++}] = \f{N_f}{3 (4\pi)^3
\varepsilon}\int d\zeta^{(-4)} ({\cal F}_V^{++})^2.
 \label{Gdiv_Nf}
 \eea
It is remarkable that the divergent contribution  \eqref{Gdiv_Nf} to
the one-loop effective action for the model \eq{Sn} is fully
determined by the external superfield $V^{++}$ in the case of the
off-shell background hypermultiplet. The expression \eqref{Gdiv_Nf}
corresponds to the action for gauge multiplet with high-derivatives,
which was introduced and studied in detail in \cite{ISZ}. To obtain
the divergent contribution to the effective action of the physical
scalar fields $\phi^i_{\cA I}$ in model \eq{Sn}, we take from the
action \eqref{Gdiv_Nf} the part, which depends only on the gauge
field $A_{M}$ \cite{ISZ}. After that one has to exclude the gauge
field $A_M$ using the classical equation of motion. We have
 \bea
\Gamma^{(1)\, N_f}_{\rm div}[\phi] = -\f{2N_f}{3 (4\pi)^3
\varepsilon}\int d^6 x (\partial^M F_{MN})^2\Big|_{A_M=-\frac{i
\epsilon_{IJ} \phi^{i \cA}_I\, \partial_{M} \phi_{i \cA J}}{\phi^{i
\cA}_I \phi_{i \cA I}}}\,, \label{Gdiv_Nf_A}
 \eea
where we denote $\partial_M = \tfrac12 (\tilde
\gamma_M)^{ab}\partial_{ab}$ and $F_{MN}$ is the abelian gauge field
strength.

\section{Concluding remarks}
In the present paper, we studied the  divergent contributions to the
one-loop effective action of the hypermultiplet model with arbitrary
self-interaction potential function in six dimensions.  We
considered in detail the model of the hypermultiplet $q^+$
interacting with the external gauge multiplet $V^{++}$ and assume
the arbitrary-gauge invariant self-coupling for hypermultiplet
\eqref{S1}.  In the six-dimensional $\cN=(1,0)$ harmonic superspace
we derived the one-loop contribution to the effective action in the
theory and calculate the divergent contribution to the effective
action \eqref{G7}. Also, we briefly discussed  the one-loop
divergent contribution to effective action \eqref{Gdiv_Nf} of model
\eqref{Sn} with $\cN_f$ number of hypermultiplets interacting with
the external gauge multiplet.

The developed above technique is a universal and can be applied to
the hypermultiplet models with rigid supersymmetry in lower
dimensions. Technically, the main distinction in these cases will be
in the calculation of the momentum and proper-time integrals in
expression \eq{G5}. The five-dimensional hypermultiplet theory does
not have one-loop logarithmic divergencies. Hence it is interesting
to study the finite contribution to the effective action. However,
in four dimensions\footnote{Many aspects of the $\cN=2$
hypermultiplet theory in four dimensions were considered in the
works \cite{Ketov1,Ketov2,Ketov3,Ketov4,Ketov5,Ketov6}.} the
one-loop effective action of the hypermultiplet with gauge-invariant
self-interacting term contains the  divergent contribution. The
developed technique in application to this model takes the result
 \bea
\Gamma^{(1)}_{4D,\, \rm div}[Q^+] = \f{1}{32\pi^2 \varepsilon}\int
d^{12}z \Psi^{--}\Psi^{++}\,. \eea For the case of potential
function $L^{+4} = \tfrac12 (\tilde Q^+Q^+)^2 $ and the on-shell
background the last expression becomes much simpler
 \bea
 \Gamma^{(1)}_{4D,\, \rm div}[Q^+] = - \f{1}{32\pi^2 \varepsilon}
 \int d^{12} z\,  \tilde Q^- Q^-\, \tilde Q^+ Q^+\,. \label{Gdiv4D}
 \eea

In the present paper, we considered the interaction of the
hypermultiplet with an abelian gauge multiplet in six dimensions. It
is natural  to generalize the above consideration to the arbitrary
non-abelian case and to study the divergent contribution to the
effective action of such model. It would be interesting to calculate
the finite contributions to the effective action of the non-abelian
hypermultiplet theory in six dimensions and consider the dimensional
reduction of such contributions to dimensions five and four. We are
going to examine these problems in the forthcoming works.

\section*{Acknowledgements}
The authors are grateful to I.L. Buchbinder for valuable
discussions. The work is partially supported by the Ministry of
Education of the Russian Federation, project No. FEWF-2020-003.


\end{document}